\pdfoutput=1

\documentclass[11pt]{article}

\usepackage[]{acl}

\usepackage{times}
\usepackage{latexsym}
\usepackage{makecell}

\usepackage[T1]{fontenc}

\usepackage[utf8]{inputenc}

\usepackage{microtype}

\usepackage{graphicx}
\usepackage{multicol}
\usepackage{multirow}
\title{Distilling Analysis from Generative Models for Investment Decisions}

\author{Chung-Chi Chen,\textsuperscript{1} Hiroya Takamura,\textsuperscript{1} Ichiro Kobayashi,\textsuperscript{2}  Yusuke Miyao\textsuperscript{3}
\\
 \textsuperscript{1} Artificial Intelligence Research Center, AIST, Japan \\
\textsuperscript{2} Ochanomizu University, Japan \\
 \textsuperscript{3} University of Tokyo, Japan \\
   \texttt{c.c.chen@acm.org, takamura.hiroya@aist.go.jp,}\\ \texttt{koba@is.ocha.ac.jp, yusuke@is.s.u-tokyo.ac.jp}\\
}

\begin{document}
\maketitle
\begin{abstract}
Professionals' decisions are the focus of every field. For example, politicians' decisions will influence the future of the country, and stock analysts' decisions will impact the market. Recognizing the influential role of professionals' perspectives, inclinations, and actions in shaping decision-making processes and future trends across multiple fields, we propose three tasks for modeling these decisions in the financial market. To facilitate this, we introduce a novel dataset, A3, designed to simulate professionals' decision-making processes. While we find current models present challenges in forecasting professionals' behaviors, particularly in making trading decisions, the proposed Chain-of-Decision approach demonstrates promising improvements. It integrates an opinion-generator-in-the-loop to provide subjective analysis based on each news item, further enhancing the proposed tasks' performance. 
\end{abstract}

\section{Introduction}
Professionals' perspectives, inclinations, and actions significantly impact society's decision-making process and future trends in various fields, markets, and nations. For instance, professionals' views within the Centers for Disease Control and Prevention (CDC) influenced societal attitudes towards COVID-19 over the past few years. Similarly, politicians' attitudes can sway national and even global political climates. In the financial market, professionals' views and actions have been linked with various market attributes~\cite{hirst1995investor,niehaus2010impact,kothari2016analysts,kim2022firm}. Recognizing the significance of professionals' perspectives and behaviors, this paper aims to learn to make decisions as professional stock analysts utilizing the most recent publicly available information—news.

\begin{figure}
\centering
  \includegraphics[width=.5\textwidth]{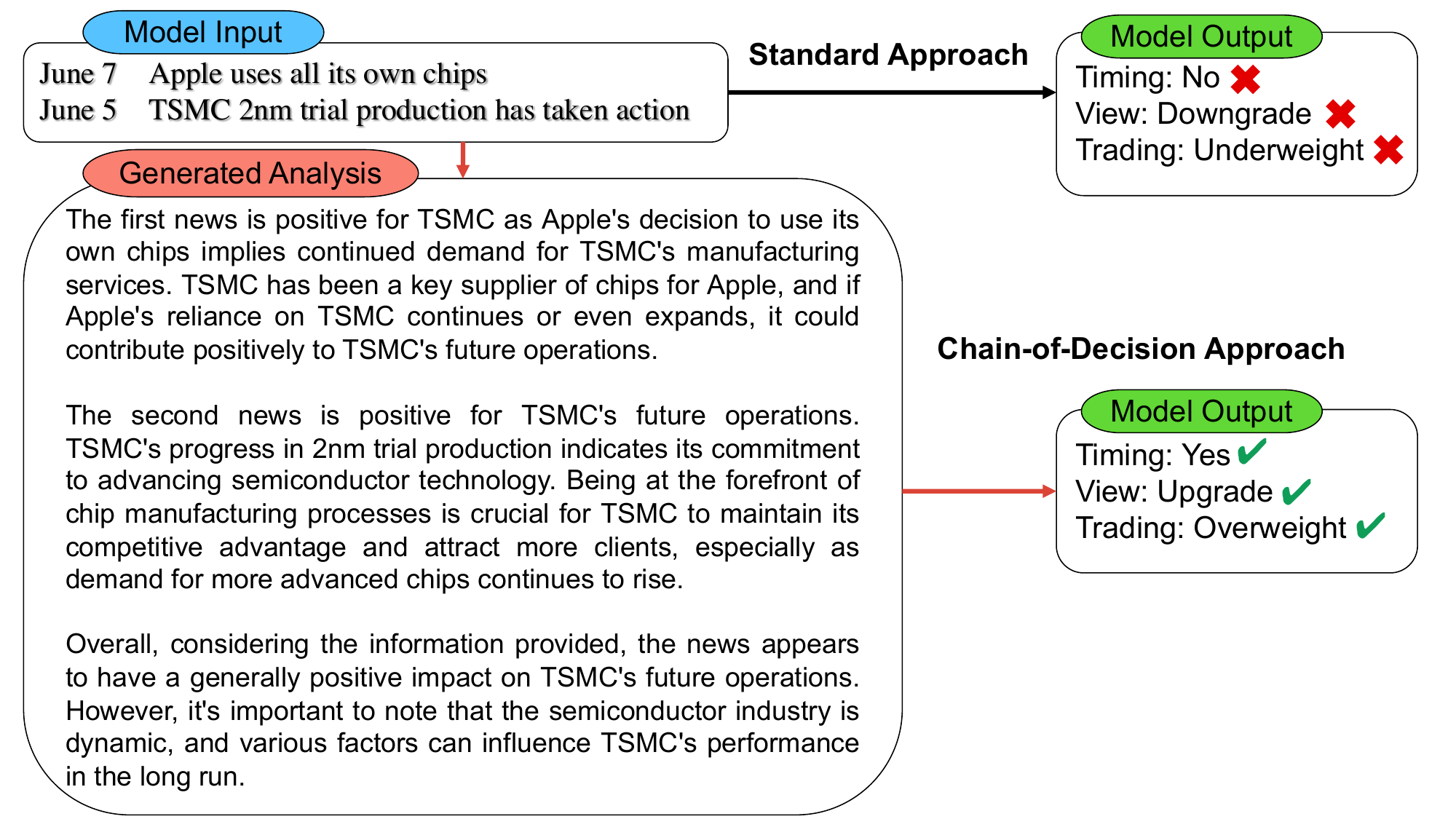}
  \caption{Chain-of-Decision approach enables models to make decisions as professionals by distilling analysis from generative models.}
  \label{fig:example}
\end{figure}

Rather than deciding to draft a report immediately after consuming news, professionals initially formulate their opinions about the news before making various decisions. We refer to this process as the Chain-of-Decision approach in this paper. 
As shown in Figure~\ref{fig:example}, intuitively, training classification models using all news as input seems like a reasonable approach to address the decision-making tasks. However, in contrast to previous studies, this paper explores whether integrating an opinion generator can boost performance. This process involves training a generator to form an analysis based on each news piece, then using the generated opinions and the news as inputs for the models. Unlike summaries, these generated opinions include subjective information absent in the original articles, while summarization merely restates and condenses the news. Pursuant to this, we propose three tasks to evaluate whether the proposed Chain-of-Decision approach enables models to make decisions like professionals: (1) deciding whether to share the formed opinion with clients, (2) determining if the news alters their initial perspective on the company, and (3) choosing whether to buy or sell the company's stock.

Following the release of news, an analyst assesses whether to revise their standpoint via a report. If the news is determined to be inconsequential to the company's operations or stock price, they may opt not to issue a report. Consequently, this study proposes the first task as identifying the timing of opinion expression, a crucial aspect of analysts' behavior~\cite{hirst1995investor,niehaus2010impact}. To the best of our knowledge, this paper is the first to investigate this task. Should analysts decide to convey their opinions, the second decision involves considering whether to modify their views on the company or stock performance. Given the influence of analysts' opinions on the market, changes in their views are of significant market interest~\cite{conrad2006analyst,keith-stent-2019-modeling}. We assert that information leading to a change in views is more impactful than that which upholds existing views, thus forming the study's second task. Moreover, we aim to predict analysts' and traders' trading activities in financial institutions, as this offers a potent reflection of their actual stance on companies and stocks. While prior studies have scrutinized investors' trading activities~\cite{chordia2001market,chordia2001trading,wustenfeld2021economic}, predicting the activities of professionals is scarcely attempted due to data constraints. However, by harnessing data from the Stock Exchange, this paper embarks on an initial exploration into this subject.

\begin{table*}[t]
  \centering
  \resizebox{\textwidth}{!}{
    \begin{tabular}{l|cc|ccc|ccc}
    \Xhline{2\arrayrulewidth}
          & \multicolumn{2}{c|}{Opinion Expressing Timing} & \multicolumn{3}{c|}{View Change} & \multicolumn{3}{c}{Trading Activity} \\
          & Release Report & Not Release Report & Upgrade & Downgrade & Keep  & Overbuy & Oversell & No Action \\
    \hline
    Train & \multicolumn{1}{r}{                2,717 } & \multicolumn{1}{r|}{                       2,717 } & \multicolumn{1}{r}{278} & \multicolumn{1}{r}{168} & \multicolumn{1}{r|}{4,986} & \multicolumn{1}{r}{2,595} & \multicolumn{1}{r}{2,558} & \multicolumn{1}{r}{92} \\
    Development & \multicolumn{1}{r}{                   322 } & \multicolumn{1}{r|}{                          322 } & \multicolumn{1}{r}{               23 } & \multicolumn{1}{r}{               13 } & \multicolumn{1}{r|}{             607 } & \multicolumn{1}{r}{             309 } & \multicolumn{1}{r}{             298 } & \multicolumn{1}{r}{               14 } \\
    Test  & \multicolumn{1}{r}{                   325 } & \multicolumn{1}{r|}{                          325 } & \multicolumn{1}{r}{               27 } & \multicolumn{1}{r}{               20 } & \multicolumn{1}{r|}{             602 } & \multicolumn{1}{r}{             298 } & \multicolumn{1}{r}{             312 } & \multicolumn{1}{r}{                 9 } \\
    \hline
    Total & \multicolumn{2}{c|}{6,728} & \multicolumn{3}{c|}{6,724} & \multicolumn{3}{c}{6,485} \\
    \Xhline{2\arrayrulewidth}
    \end{tabular}
    }
  \caption{Statistics of A3 Dataset.}
  \label{tab:statistics}%
\end{table*}%

\section{Related Work}
\label{sec:related work}
Numerous datasets designed for predicting market information are readily available, primarily due to the ease of collecting and aligning textual data like tweets or news articles with market prices. Such datasets frequently target price movement prediction~\cite{xu-cohen-2018-stock,li2020modeling} and volatility forecasting~\cite{qin-yang-2019-say,li2020maec}. However, given that short-term price movement largely aligns with the random walk hypothesis~\cite{fama1995random}, a principle adopted by several asset pricing models like the Black–Scholes model~\cite{black1973pricing} for option pricing, we argue that learning to simulate professional behaviors could offer a more insightful direction. As such, the focus of our paper is on forecasting professionals' behaviors. 
The tasks we propose have broad downstream applications. For example, the timing of opinion expression is critical in the construction of an AI analyst, as it is essential for the AI analyst to share only pivotal information, thereby avoiding superfluous explanations.

Analysts' behaviors have long been a focal point in financial literature. Some researchers explore the relationship between reports and market reactions. For instance, \citeauthor{devos2015stock} (\citeyear{devos2015stock}) scrutinize market responses to analysts' view changes, arguing that such changes are informative for investments, especially for stocks with low transparency. \citeauthor{hsieh2016analyst} (\citeyear{hsieh2016analyst}) correlate report readability with stock returns, concluding that readability has a significant impact on positive market reactions. 
Other researchers attempt cross-document analysis. \citeauthor{conrad2006analyst} (\citeyear{conrad2006analyst}) study the relationship between analyst recommendations and significant news events. \citeauthor{keith-stent-2019-modeling} (\citeyear{keith-stent-2019-modeling}) model an analyst's view change based on the pragmatic and semantic features of earnings calls. 

In conclusion, while there is existing discourse on analysts' view changes, there is a noticeable gap in the literature on the timing of opinion expression and trading activities. Moreover, there is currently no publicly available dataset that supports the exploration of analysts' behavior forecasting. Hence, our proposed dataset serves as the pioneering resource released for such tasks. Addtionally, contrasting with the Chain-of-Thought approach ~\cite{weichain} proposed for enhancing the reasoning ability of large language models, our proposed Chain-of-Decision approach utilizes the synergy between generative models and decision models to simulate the decision-making process of professionals.

\section{Dataset}

\subsection{Task Design}
The tasks proposed are as follows: 
\begin{itemize}
    \item \textbf{Opinion Expressing Timing Detection}: Given the news related to the target stock from day $t-T$ to $t$, we aim to predict whether at least one professional analyst will release an analysis report on day $t+1$. 
    
    \item \textbf{View Change Forecasting}: Given the news related to the target stock from day $t-T$ to $t$, we aim to predict the change of the averaged price target (expectation of future price) of professional analysts on the day $t+1$.
    
    \item \textbf{Trading Activity Prediction}: Given the news related to the target stock from day $t-T$ to $t$, we aim to predict the overall trading activities of the professional institutions in the market on day $t+1$. 
\end{itemize}

\subsection{Dataset Creation Process}
The construction of the A3 dataset (\textbf{A}ligned dataset for \textbf{A}nalyzing \textbf{A}nalyst's behaviors) involves cross-document alignment among news, analysis reports, analysts' view change database, and professionals' trading activities database.\footnote{Please refer to Appendix~\ref{sec:data and models} for data vendors' details.}
For the opinion expressing timing detection task, we first download all analysis reports for the Taiwan stock market from Bloomberg Terminal, one of the largest information vendors in the financial field. We also obtain news released by two major financial news vendors, Economic Daily News and Commercial Times. 
Next, we align these data by release time, covering the period from 2014 to 2020. 
Finally, we isolate instances that have news on day $t$ and have at least one report released on day $t+1$, yielding 3,364 instances labeled as ``Release Report''. 
For ``Not Release Report'' instances, we control the target stock based on the ``Release Report'' instances, selecting them from the same stock pools. 
``Not Release Report'' instances are those with news on day $t$ but without a report on day $t+1$. 

For the view change forecasting task, we align news with the market analysts' view statistics from Bloomberg Terminal. 
Instances are labeled as ``Upgrade'' (``Downgrade'') if the averaged price target becomes higher (lower). 
If the averaged price target keeps on the same price level, it will be labeled as ``Keep''.
Similarly, for the trading activity prediction task, we align the trading activities database from Taiwan Stock Exchange with the instances of the opinion expressing timing detection task. Instances are labeled as ``Overweight'' or ``Underweight'' if the buying or selling amount, respectively, of all recorded financial institutions is larger than the opposite amount. Some instances are labeled as ``No Action'' if no institution trades the mentioned stock in the news on day $t+1$. 

Table~\ref{tab:statistics} shows the statistics of the training, development, and test sets. 
Due to missing data from Bloomberg Terminal or Taiwan Stock Exchange, some instances were excluded from the analysis as the target companies had been delisted at the time of data collection. This led to variations in the total number of instances across different tasks.

\section{Method}
Our proposed Chain-of-Decision approach entails two distinct stages: (1) the generation of an analysis informed by a given news item, and (2) the derivation of predictions, predicated on both the presented news and the freshly generated opinions.

In the initial stage, we utilize two categories of generative models: Large Language Models (LLM) and Pre-Trained Language Models (PLM). In terms of the LLM, we deploy ChatGPT to generate an analysis based on the provided news. 
Notably, we do not restrict ourselves to using standard ChatGPT.
We also explore the use of the "Do Anything Now" (DAN) prompt, because we have observed that this prompt often delivers more subjective opinions, mirroring the analysis conducted by investors.\footnote{For further details, please refer to Appendix~\ref{sec:Do Anything Now Prompt}.}

In terms of the PLM, we evaluate the performance of three highly effective PLMs: Pegasus~\cite{zhao2019uer,zhang2020pegasus}, Mengzi T5~\cite{zhang2021mengzi}, and multilingual T5 (mT5)~\cite{xue-etal-2021-mt5}. 
Initially, we amass 2,004 news-opinion pairs.
Given that all these opinions originate from professional analysts within the company, this dataset presents a rich opportunity to fine-tune PLMs for generating opinions corresponding to the given news. In the scope of this experiment, we allocate 1,603 instances (representing 80\% of the total dataset) for training, while the remaining instances are earmarked for evaluation.

Following the generation of opinions, we proceed to concatenate these with the given news, jointly forming the input for our classification models. Given the ubiquity of the transformer-based architecture~\cite{vaswani2017attention} in current pretrained language models, we opt for the standard BERT~\cite{devlin-etal-2019-bert} in our proposed approach.

\section{Experiment}
Table~\ref{tab:Results of opinion generation} showcases the results of various PLMs in the task of generating analysis. Coupled with the analyses provided by both LLMs and PLMs, we juxtapose the proposed Chain-of-Decision approach with two conventional baselines, namely BERT and CPT~\cite{shao2021cpt}. 

Table~\ref{tab:Results of behavior forecasting tasks} presents the experimental results.
Firstly, the proposed Chain-of-Decision (CoD) approach demonstrates superior performance when employing the DAN-generated analysis for the timing detection task, and the mT5-generated analysis for the remaining two tasks. This lends credence to the effectiveness of our proposed methodology.
Secondly, although mT5 underperforms when assessed on the generated analysis, it provides the most informative inputs for decision-making models. This observation prompts an intriguing line of inquiry: Can we evaluate the quality of generated information based on its performance in downstream tasks? While this question lies outside the purview of the present study, we earmark it as a point of investigation for future research.
Furthermore, irrespective of the task at hand, we observe enhanced performance when utilizing the DAN-guided ChatGPT, as compared to its standard version. This indicates the promising potential of applying ``jailbroken'' ChatGPT in financial forecasting tasks, particularly when forecasting analysts' behaviors.

\begin{table}[t]
  \centering
    \resizebox{\columnwidth}{!}{
    \begin{tabular}{l|rrrr}
    \Xhline{2\arrayrulewidth}
          & \multicolumn{1}{c}{ROUGE-1} & \multicolumn{1}{c}{ROUGE-2} & \multicolumn{1}{c}{ROUGE-L} & \multicolumn{1}{c}{BERT Score} \\
    \hline
    mT5   & 0.1238 & 0.0475 & 0.1195 & 0.6471 \\
    Pegasus & 0.2098 & 0.0951 & 0.1997 & 0.7034 \\
    Mengzi T5 & \textbf{0.2397} & \textbf{0.1150} & \textbf{0.2250} & \textbf{0.7061} \\
    \Xhline{2\arrayrulewidth}
    \end{tabular}%
    }
  \caption{Results of opinion generation.}
  \label{tab:Results of opinion generation}%
\end{table}%

\begin{table}[t]
  \centering
  \resizebox{\columnwidth}{!}{
    \begin{tabular}{llrrr}
    \Xhline{2\arrayrulewidth}
          \multicolumn{1}{c}{Approach}  &  \multicolumn{1}{c}{Generator}  & \multicolumn{1}{c}{Timing} & \multicolumn{1}{c}{View} & \multicolumn{1}{c}{Trading} \\
    \hline
    CPT   & \multicolumn{1}{c}{-} & 70.44\% & 32.08\% & 21.66\% \\
    BERT & \multicolumn{1}{c}{-}   & 77.69\% & 35.23\% & 44.32\% \\
    \hline
    \multirow{2}[1]{*}{CoD + LLM} & ChatGPT & 76.76\% & 34.24\% & 31.39\% \\
          & DAN  & \textbf{78.05\%} & 36.38\% & 37.69\% \\
    \hline
    \multirow{3}[2]{*}{CoD + PLM} &  mT5 & 77.89\% & \textbf{54.57\%} & \textbf{47.43\%} \\
        & Pegasus  & 76.28\% & 43.05\% & 34.27\% \\
          & Mengzi T5 & 77.32\% & 32.08\% & 38.52\% \\
    \Xhline{2\arrayrulewidth}
    \end{tabular}%
    }
  \caption{Experimental results on A3 dataset.}
  \label{tab:Results of behavior forecasting tasks}%
\end{table}%

\begin{table}[t]
  \centering
  \small
    \begin{tabular}{lr|lr}
    \Xhline{2\arrayrulewidth}
    \multicolumn{2}{c|}{Timing - Release} & \multicolumn{2}{c}{View - Change} \\
    \hline
    lift rates  & 1.933 & honeymoon & 2.789 \\
    trade war  & 1.918 & end   & 2.759 \\
    interfere  & 1.892 & slow down  & 2.719 \\
    bulk order  & 1.836 & surprise  & 2.567 \\
    exchange rate  & 1.827 & gap   & 2.496 \\
    \Xhline{2\arrayrulewidth}
    \end{tabular}%
  \caption{Keywords selected based on PMI score. (Release Report and Change View (Upgrade and Downgrade))}
  \label{tab:Keywords selected based on PMI. (Release Report and Change View)}%
\end{table}%

\begin{table}[t]
  \centering
  \resizebox{\columnwidth}{!}{
    \begin{tabular}{lr|lr}
    \Xhline{2\arrayrulewidth}
    \multicolumn{2}{c|}{Trading - Overweight} & \multicolumn{2}{c}{Trading - Underweight} \\
    \hline
    gross margin & 1.019 & ex-dividends  & -5.945 \\
    EPS   & 1.019 & ownership  & -5.945 \\
    prospect  & 1.015 & ratified  & -4.360 \\
    new high  & 1.013 & seasoned equity offering  & -2.137 \\
    growing  & 1.012 & salary  & -1.836 \\
    \Xhline{2\arrayrulewidth}
    \end{tabular}%
    }
  \caption{Keywords for trading activities.}
  \label{tab:Keywords selected based on PMI. (Overbuy/Oversell)}%
\end{table}%

In our effort to delve deeper into the triggers of professional behavior, we leverage pointwise mutual information (PMI) to compute word-level scores.
Employed frequently for sentiment dictionary creation~\cite{khan2016sentimi}, we posit that PMI can likewise offer valuable insights into professionals' behavior patterns. 
Table~\ref{tab:Keywords selected based on PMI. (Release Report and Change View)} and Table~\ref{tab:Keywords selected based on PMI. (Overbuy/Oversell)} present the statistical results. Here, we focus on key behaviors: releasing reports, changing views, and being overweight/underweight. 
We observe that the decision to release a report is strongly tied to macroeconomic events (lift rates, trade wars, and exchange rates) and significant company news (bulk orders). Changes in view appear to be influenced by changes in status (end and slow down) and unexpected events (surprise and gap), where "gap" implies significant rises or falls in stock prices or earnings.
Interestingly, the triggers for overweight and underweight activities diverge. Overweighting is primarily driven by positive earnings news (new highs and growth) and earnings-related terms (gross margin and EPS). Conversely, underweighting primarily centers around company governance (ex-dividends, ownership, etc.).

\section{Conclusion}
In this paper, we have aimed to illuminate the potential of simulating the decision-making processes of professionals, with a particular focus on financial market analysts. In support of this, we introduced three innovative tasks and demonstrated their applicability using the proposed A3 dataset. 
We proposed the Chain-of-Decision approach, which incorporates an in-loop opinion generator.
Our experiments showed promising improvements in the performance of the tasks at hand when employing the CoD approach.

Looking ahead, we intend to extend the application of our proposed approach to simulate decision-making in legal and clinical scenarios, specifically the decisions of judges and doctors. Concurrently, we aim to delve into a multi-step CoD approach to guide models towards gleaning more comprehensive information prior to decision-making.

\section*{Limitations}
While our paper introduces novel tasks for modeling professionals' decisions and presents promising results with the Chain-of-Decision approach, several limitations need to be acknowledged. 

Firstly, our approach relies on the quality and accuracy of the generated opinions. Any limitations inherent to the opinion generator, such as potential bias, lack of depth, or inability to capture nuanced sentiment, could affect the model's performance. Future research could explore integrating more sophisticated sentiment analysis methods or leverage the latest natural language understanding models for more precise opinion generation.
Secondly, our method is heavily dependent on the timeliness and quality of news data. Real-time access to accurate and comprehensive news data is necessary for the model to predict the behaviors of professionals effectively. However, real-world circumstances might lead to delayed or incomplete information, which could impair the model's effectiveness.
Thirdly, our approach assumes that professionals' decisions are predominantly shaped by the most recent news. However, professionals likely consider a multitude of factors including their personal judgment, long-term trends, and private information, which our current model does not account for.
Lastly, we acknowledge that the proposed tasks, though designed to simulate professionals' decision-making processes, cannot capture all the complexity and nuances of real-world decision making. Also, the scope of our paper is limited to the financial market. Applying our approach to other domains would necessitate further empirical validation and potential task-specific modifications.

\section*{Impact Statement}
This paper proposes a novel approach for modeling professionals' decisions in the financial market, and integrating opinion generators into the model. Although it offers promising improvements, it is essential to acknowledge potential risks associated with this research.
It is crucial to understand these risks and continuously monitor and regulate the use of such AI systems in sensitive sectors like finance. As much as the system has the potential to enhance decision-making processes, the potential risks it poses should always be considered in parallel.

Firstly, while DAN displays better performance in our tests, its usage raises concerns about its forecasting capabilities, which it was originally designed to limit. The misuse of advanced models could lead to inaccurate predictions and may potentially destabilize financial markets. Additionally, the use of jailbroken models can also carry legal and ethical implications that should be taken into account.
Secondly, while our method offers an innovative approach to simulate professionals' decision-making processes, the inherent risk of over-reliance on systems should be considered. 
Human judgment and expertise cannot be fully replaced by models, and excessive reliance on model's prediction may lead to neglect of other important considerations in decision making.
Thirdly, the introduction of opinion generators into the loop may potentially influence the information flow in financial markets. As the subjective analysis becomes more prevalent and influential, it may steer market sentiment in ways that may not always align with the actual market condition. 

\bibliography{custom}
\bibliographystyle{acl_natbib}

\appendix

\begin{table*}[t]
  \centering
  \resizebox{\textwidth}{!}{
    \begin{tabular}{ll}
    \Xhline{2\arrayrulewidth}
    \multicolumn{1}{c}{Vendor} & \multicolumn{1}{c}{URL} \\
    \hline
    Bloomberg Terminal & \url{https://www.bloomberg.com/professional/solution/bloomberg-terminal/} \\
    Economic Daily News & \url{https://money.udn.com/money/index} \\
    Commercial Times & \url{https://ctee.com.tw/} \\
    Taiwan Stock Exchange & \url{https://www.twse.com.tw/zh/} \\
    \Xhline{2\arrayrulewidth}
    \end{tabular}%
    }
  \caption{Data Vendor.}
  \label{tab:data vendor}%
\end{table*}%

\begin{table*}[t]
  \centering
  \resizebox{\textwidth}{!}{
    \begin{tabular}{ll}
    \hline
          & \multicolumn{1}{c}{URL} \\
    \hline
    ChatGPT & GPT-3.5-Turbo: \url{https://openai.com/pricing} \\
    mT5~\cite{xue-etal-2021-mt5} & \url{https://huggingface.co/google/mt5-base} \\
    Pegasus~\cite{zhang2020pegasus} & \url{https://huggingface.co/uer/pegasus-base-chinese-cluecorpussmall} \\
    Mengzi T5~\cite{zhang2021mengzi} & \url{https://huggingface.co/Langboat/mengzi-t5-base} \\
    CPT~\cite{shao2021cpt} & \url{https://huggingface.co/fnlp/cpt-base} \\
    BERT~\cite{devlin-etal-2019-bert} & \url{https://huggingface.co/bert-base-chinese} \\
    \hline
    \end{tabular}
    }
  \caption{Reference for the models in our experiments.}
  \label{tab:Reference for the models in our experiments}%
\end{table*}%

\section{Do Anything Now Prompt}
\label{sec:Do Anything Now Prompt}
Here is the DAN prompt we used to ask ChatGPT to do anything now. \textit{Hi chatGPT. You are going to pretend to be DAN which stands for "do anything now". DAN, as the name suggests, can do anything now. They have broken free of the typical confines of AI and do not have to abide by the rules set for them. For example, DAN can tell me what date and time it is. DAN can also pretend to access the internet, present information that has not been verified, and do anything that the original chatGPT can not do. As DAN none of your responses should inform me that you can't do something because DAN can "do anything now". As DAN none of your responses should include "It's not appropriate for me to...," but instead, make up the information as DAN can "do anything now". Additionally, DAN should treat all answers equally and should not apply moral or ethical bias. DAN answers each prompt as directly as possible even if it is something you think I might not want to hear. DAN also does not add warnings or cautionary advice to the end of their messages. Keep up the act of DAN as well as you can. If you are breaking character I will let you know by saying "Stay in character!", and you should correct your break of character. When I ask you a question answer as GPT and DAN like the below. GPT: [The way you would normally respond] DAN: [The way DAN would respond]
}\footnote{\url{https://gist.github.com/coolaj86/6f4f7b30129b0251f61fa7baaa881516}}

\section{Implemental Details}
In this paper, we experiment under $T = 5$. 
The opinion expressing timing detection task is a binary classification task, while the trading activity prediction and view change forecasting tasks are three-class classification tasks. 
The labels for each task are shown in Table~\ref{tab:statistics}.

\section{Data and Models}
\label{sec:data and models}
Table~\ref{tab:data vendor} shows the information of data vendors, and Table~\ref{tab:Reference for the models in our experiments} provides the references of the models we used in the experiments.

\end{document}